\documentclass[article,aps,twocolumn,prl,letterpaper,superscriptaddress,showpacs,nofootinbib]{revtex4}
\usepackage{amsmath,amssymb}
\usepackage{color}
\usepackage[pdftex]{graphicx} 
\usepackage{hyperref}

\begin{document}
\newcommand{\comment}[1]{}
\newcommand{\E}{\mathrm{E}}
\newcommand{\Var}{\mathrm{Var}}
\newcommand{\bra}[1]{\langle #1|}
\newcommand{\ket}[1]{|#1\rangle}
\newcommand{\braket}[2]{\langle #1|#2 \rangle}
\newcommand{\be}{\begin{equation}}
\newcommand{\ee}{\end{equation}}
\newcommand{\ba}{\begin{eqnarray}}
\newcommand{\ea}{\end{eqnarray}}
\newcommand{\SD}[1]{{\color{magenta}#1}}
\newcommand{\HME}[1]{{\color{green}#1}}
\newcommand{\rem}[1]{{\color{cyan} \sout{#1}}}
\newcommand{\alert}[1]{\textbf{\color{red} \uwave{#1}}}
\newcommand{\Y}[1]{\textcolor{BurntOrange}{#1}}
\newcommand{\R}[1]{\textcolor{red}{#1}}
\newcommand{\B}[1]{\textcolor{blue}{#1}}

\title{Classical demonstration of frequency dependent noise ellipse rotation using Optomechanically Induced Transparency}

\author{Jiayi Qin}
\email{jiayiqinphysics@gmail.com}
\affiliation{School of Physics, University of Western Australia, WA 6009, Australia}
\author{Chunnong Zhao}
\email{chunnong.zhao@uwa.edu.au}
\affiliation{School of Physics, University of Western Australia, WA 6009, Australia}
\author{Yiqiu Ma}
\email{myqphy@gmail.com}
\affiliation{School of Physics, University of Western Australia, WA 6009, Australia}
\author{Xu Chen}
\affiliation{School of Physics, University of Western Australia, WA 6009, Australia}
\author{Li Ju}
\affiliation{School of Physics, University of Western Australia, WA 6009, Australia}
\author{David G. Blair}
\affiliation{School of Physics, University of Western Australia, WA 6009, Australia}

\begin{abstract}
Cavities with extremely narrow linewidth of $10-100$ Hz are required for realizing frequency dependent squeezing to enable gravitational wave detectors to surpass the free mass standard quantum limit over a broad frequency range. Hundred-meter-scale high finesse cavities have been proposed for this purpose. Optomechanically induced transparency (OMIT) enables the creation of  optomechanical cavities in which the linewidth limit is set by the extremely narrow linewidth of a high $Q-$factor mechanical resonator. Using an 85mm OMIT cavity with a silicon nitride membrane, we demonstrate a tunable linewidth from 3Hz up to several hundred Hz and frequency dependent noise ellipse rotation using classical light with squeezed added noise to simulate quantum squeezed light. The frequency dependent noise ellipse angle is rotated in close agreement with predictions.
\end{abstract}

\maketitle


{\it Introduction---}
The coherent interaction of laser radiation with widely spaced mirror test masses is used to measure gravitational wave induced motions in interferometric gravitational-wave detectors. The sensitivity of first generation gravitational wave (GW) detectors such as LIGO reached the quantum shot noise limit in the high frequency part of the spectrum. In the second generation detectors now under construction,  quantum radiation-pressure noise is expected to dominate at low frequencies, while shot noise  will  dominate at high frequencies. A region around 100Hz is limited by classical test mass thermal noise, but as better optical coatings and test masses become available, future detectors should be limited mostly by quantum noise.

In the late 1960s, Braginsky pointed out that there exists a Standard Quantum Limit (SQL) in gravitational wave detectors \cite{Braginsky1}, and proposed that quantum non-demolition (QND) devices could beat the SQL \cite{Braginsky2}. In 2001, Kimble et al. \cite{FDKimble} proposed  QND interferometer designs that involved the use of pairs of successive filter cavities for realizing frequency-dependent squeezing (FDS) of the input squeezed light, or frequency dependent (FD) homodyne detection in which the output field of the detector is filtered in the frequency dependent way. They pointed out that pairs of successive Fabry-P\'{e}rot filter cavities can be used to convert ordinary squeezed light into FD squeezed light such that the sensitivity of the detector across the entire frequency band is improved below the SQL. Similar cavities could also be installed between the interferometer output and the ordinary homodyne detection to realize FD-homodyne detection. Recently Chelkowski et al. demonstrated FD squeezed vacuum using a short filter cavity in the MHz range \cite{Squeezerotate}. In 2012, Stefszky et al. demonstrated 11.6 dB squeezing in aLIGO detection band \cite{squeezegeneration}.

To match the filter cavity linewidth to the corner frequency of ground based laser interferometers where the shot noise becomes higher than the radiation pressure noise, the filter cavity must meet very demanding specifications that require very long optical cavities with very low optical losses. To optimize the sensitivity, adjustable cavity linewidth and offset frequency locking are also required. For example, an aLIGO type GW detector requires hundred-meter scale filter cavities of linewidth of $\sim$100Hz to optimize sensitivity \cite{loss}.

In order to realize FDS in tabletop filter cavities, Ma et al. proposed using optomechanically induced transparency (OMIT) effect to achieve  tunable narrow linewidth \cite{Yiqiu}. The idea of OMIT is analogous to the electromagnetically induced transparency (EIT) phenomena discovered in three-level atomic systems \cite{EIT1}. This phenomenon was widely recognized and applied in various fields of optics \cite{EITreview, EITprogress, EITreview2, PropapationDynamics}. Recently, Mikhailov et al. proposed to use EIT  to create FD squeezed vacuum for GW detectors. However, the high optical loss of EIT is still an issue \cite{squeesevacuum}.

The OMIT phenomenon has been studied and demonstrated by various research groups. Weis et. al \cite{OMITweis} presented OMIT phenomena in a toroidal microcavity and achieved a tunable linewidth of 50$\sim$500 kHz compared with the 15 MHz  linewidth of the optical mode. Recently, Karuza et.al\cite{Karuza} demonstrated the OMIT effect in a membrane-in-the-middle setup at room temperature. They reported a maximum signal time advance $\tau^{T}\approx-108$ ms of a probe pulse, which implies an OMIT linewidth  much narrower than those mentioned above \cite{OMITweis}.

In reference \cite{Yiqiu}, Ma et al. theoretically investigated using optomechanical interactions to achieve  tunable narrow linewidth in a  tabletop filter cavity which could in principle be used to convert ordinary squeezed vacuum to FD squeezed vacuum, or to shift the local oscillator light phase with the appropriate frequency dependence in output homodyne detection scheme. This configuration allows possible realization of FDS in a table-top experiment.

The experimental challenges of such a device are the stringent demands for very low temperature and a very low-loss optical cavity if optical dilution were used. The optical dilution is realized by using optomechanical interactions to increase the effective Q$-$factor of a given mechanical resonator. To evade quantum back-action of strong optical dilution, Chang et al. \cite{chang} and Ni et al. \cite{Ni} proposed using a nonlinear quadratic optical trap, and more recently Korth et al. \cite{korth} proposed detecting quantum back-action in the outgoing field and actively feeding back to the system. The requirement for a low-temperature operation of such a device is \cite{Yiqiu}:
\be
 \frac{8 k_B T}{Q_m}<\hbar\Gamma_{\rm opt},
\ee
where $T$ is the environmental temperature, $Q_m$ is the mechanical $Q-$factor and $\Gamma_{\rm opt}$ is the effective cavity linewidth. For an OMIT cavity with $\Gamma_{\rm opt}\approx2\pi\times 100$ Hz, the temperature requirement is $T/Q_m<6.0\times10^{-10}$ K. In 2008, Zwickl et al. reported that silicon nitride membrane has mechanical $Q>10^6$ at 293 K and $Q>10^7$ at 300 mK \cite{highqMembrane}. And recently, Jayich et al. observed a mechanical quality factor $Q>4\times 10^6$ of a silicon nitride membrane placed at the center of an optical cavity at 400 mK \cite{cryo}.

In this letter, we use a noise-added signal light to mimic the squeezed light in a room temperature system in which a control light is injected into the same port for generating OMIT effect. We demonstrate frequency dependent noise ellipse rotation in a tunable OMIT cavity in which the linewidth can be tuned from 3Hz to several hundred Hz. In order to measure the noise ellipse rotation of the single-mode signal field by the lock-in technique \cite{EIT}, we detect the beating of the signal field and the control field at the transmission port of the coupled system. The result proves that the OMIT cavity has the same amplitude and phase response as a simple filter cavity, which can rotate the noise ellipse of a classical signal light with squeezed added noise in close agreement with the theoretical phase response. This shows the potential of FD squeezed vacuum generation in a small scale compact system with future implementations of low temperature
 environment and proper optical dilution.

The core elements of our OMIT apparatus consists of an 85 mm high-finesse optical cavity with a high stress silicon nitride membrane, which has a  quality factor of $\sim1.5 \times10^6$ at the mechanical resonance. By changing the frequency separation between signal field and control field, we  observed the angle rotation of noise ellipses of the signal light, which is shown in Fig. \ref{rotate}.


\begin{figure}[!t]
\includegraphics[width=0.42\textwidth]{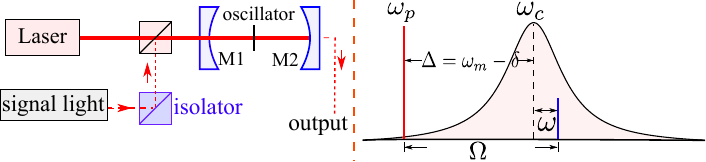}
\caption{Configuration schematics (left) and frequency relationships. The signal light with squeezed added noise having frequency $\omega$ respect to cavity resonance $\omega_c$ is injected into an optical cavity with a high Q$-$factor membrane in the middle which acts as an oscillator at the resonant frequency $\omega_{m}$. The position of the membrane is chosen to introduce a linear optomechanical coupling. The radiation
pressure from the beating between the signal light $\omega_{s}=\omega_{p}+\Omega$ and the control laser $\omega_p$ coherently
drives the mechanical oscillator which in turns creates sidebands destructively interfering with the signal
light, which in effect rotates the noise ellipse angle.}
\label{fig2}
\end{figure}

 {\it Introduction---} Optomechanical interaction happens when beating between the control and signal optical fields creates a radiation pressure that induces mechanical motion. This mechanical motion then modulates the control field to produce an upper sideband which has the same frequency as the signal field, but of opposite sign. That is, the optomechanical interaction causes a backaction that reduces the signal field to generate the OMIT effect.
 The linewidth of an OMIT cavity is determined by the sum of the mechanical damping and the optomechanical damping associated with laser-cooling of the membrane by a red-detuned control field. In the strong optomechanical coupling regime, this tunable optomechanical damping dominates over mechanical damping and thus sets the linewidth of the system.

As shown in the right of Fig.\ref{fig2}, the control laser at frequency $\omega_{p}$ maintains a strong control field $\overline {a}_{p}e^{-i\omega_{p}t}$ in the cavity at the resonant frequency $\omega_c$. A weak signal light is injected into the same port (M1) as a small input term $\delta \hat a_{\rm in}=\hat{a}^s_{\rm in} e^{-i\omega_{s} t}$. The frequency difference $\Omega$ between the control field $(\omega_{p})$ and signal field $(\omega_{s}=\omega_{c}+\omega)$ needs to be close to the mechanical resonant frequency $\omega_{m}$ for efficient
driving of the mechanical mode.

The Hamiltonian which describes the system here is given by:
\be
\hat{H}=\hbar(\omega_{c}+G_{0}\hat{x}) \hat{a}^{\dag}\hat{a}+\hat{H}_{m}+\hat{H}_{\gamma}.
\label{hamiltonian}
\ee
Here, $\hat{H}_{m}=\hat{p}^{2}/2 m+m\omega_{m}^{2}\hat{x}^{2}/2$  is the Hamiltonian of the mechanical oscillator. $H_\gamma=-i\hbar\sqrt{2\gamma_1}\hat{a}\hat{a}_{in}^{\dagger}
-i\hbar\sqrt{2\gamma_2}\hat a\hat b_{in}^{\dagger}+h.c$ describes the interaction between the intra-cavity field $\hat{a}$ and external electromagnetic fields $\hat{a}_{\rm in}$ and $\hat b_{\rm in}$ with interaction strengthes $\gamma_1=cT_1/4L$ and $\gamma_2=cT_2/4L$ through the cavity mirrors M1 and M2 respectively. $G_{0}$ is the linear optomechanical coupling strength \cite{coupling, finesse}.

 Since the signal light $\hat a^s_{\rm in}$ in our experiment is a classical field injected from the left in Fig. \ref{fig2}, we neglect the vacuum fluctuation. In the rotating frame at frequency $\omega_{p}$, we have:
\begin{subequations}
\begin{align}
 \chi(\Omega)\hat{x}(\Omega)=-\hbar \bar{G}_{0}[\hat{a}(\Omega)+ \hat a^{\dagger}(-\Omega)]+\hat{F}_{\rm th},\label{eq eom1}\\
\hat{a}(\Omega)=\frac{\bar{G}_{0}\hat{x}(\Omega)}{\Omega-\Delta+i\gamma}+ \frac{i\sqrt{2\gamma_1}\hat{a}^s_{\rm in}(\Omega)}{\Omega-\Delta+i\gamma},\label{eq eom2}
\end{align}
\end{subequations}
where $\gamma_{m}$ and $\gamma=\gamma_1+\gamma_2$ are the linewidths of the mechanical oscillator and the cavity, $\chi(\Omega)=m(\omega^2_m-\Omega^2-i\gamma_m\Omega)$ is the mechanical response function, and $\bar{G}_0$ is defined as $G_{0}\bar{a}$. The thermal force is a random force of which the spectral density is given by $S^{\rm th}_{\rm FF}(\omega)=4m\gamma_mk_BT$ \cite{thermal}. $\Delta=\omega_c-\omega_p$ represents the detuning of the control beam to the cavity resonance and we choose $\Delta=\omega_m-\delta\sim\omega_{m}$.

Since the lower sideband of the cavity mode is far detuned from the cavity resonance point, our system satisfies the resolved sideband limit. In this case, we only have the upper sideband cavity mode:
\be\label{upsideband}
\hat a(\Omega)\approx\frac{\bar{G}_0}{i\gamma}\hat x(\Omega)+\frac{\sqrt{2\gamma_1}}{\gamma}\hat a^s_{\rm in}(\Omega).
\ee
Here we make use of the near resonance approximation: $\Omega-\Delta=\omega\ll\gamma$. Substitute \eqref{upsideband} into the radiation
pressure force term in ~\eqref{eq eom1}, the equation of motion for mechanical oscillator can be written as:
\be\label{eff_eom}
\chi_{\rm eff}(\omega)\hat{x}(\Omega)=-\frac{\sqrt{2\gamma_1}\hbar\bar{G}_0}{\gamma}\hat a^s_{\rm in}(\Omega)+\hat{F}_{\rm th},
\ee
in which the effective mechanical response function $\chi_{\rm eff}$ is: $\chi_{\rm eff}(\omega)\approx-2m\omega_m(\omega-\delta+i\gamma_m)-i\hbar\bar{G}_0^2/\gamma$. Here, we use the near-resonance approximation for $\chi_{\rm eff}(\Omega)$ in \eqref{eq eom1}. In our system, the thermal force $\hat{F}_{\rm th}$ with the spectrum density $S^{\rm th}_{\rm FF}\sim10^{-30}$ $\rm N^2/\rm Hz$ is negligible compared with the radiation pressure force of the classical added noise field and the control field with the spectrum density $S^{n}_{\rm FF}\approx2\gamma_1(\hbar\bar{G}_0|\delta \hat a^s_{\rm in}|)^2/\gamma^2\gamma_m\sim 10^{-21}$ $\rm N^2 /\rm Hz$ for measurements shown in Fig. \ref{phaserotation} and \ref{rotate}.

Substitute \eqref{eff_eom} into \eqref{upsideband} and use the relation
between the transmitted field and injected field $\hat a_{\rm trans}=\sqrt{2\gamma_2}\hat a$,
we have effective transmissivity:
\be\label{t}
t(\Omega)=2\sqrt{\eta_c(1-\eta_c)}\frac{\Omega-\omega_{m}+i\gamma_{m}}{\Omega-\omega_{m}+i\gamma_m+i\Gamma_{\rm opt}},
\ee
where the cavity coupling parameter is $\eta_c=\gamma_1/(\gamma_1+\gamma_2)$.

 The characteristic frequency $\Gamma_{\rm opt}$ is defined to be equal to $\hbar \bar{G}^2_0/2m\omega_m\gamma$. Equation \eqref{t} shows that OMIT can give rise to a filter cavity with linewidth that can approach $\gamma_{m}$. Using a mechanical oscillator with a sufficient high-$Q$ factor, we can create an OMIT cavity with linewidth of several hundred Hz in which $\gamma_m$ is negligible since $\gamma_m\ll\Gamma_{\rm opt}$. When the frequency detuning $|\Omega-\omega_m|\gg\sqrt{\gamma_m\Gamma_{\rm opt}}$, the phase $\phi(\Omega)$ of the system transmissivity can be written as:
\be
\phi (\Omega)=-\arctan{(\frac{\Gamma_{\rm opt}}{\Omega-\omega_{m}})}
\ee
 which is equivalent to the transmissivity phase response of a simple Fabry-P$\acute{e}$rot cavity with the resonant point at $\Omega=\omega_m$. The linewidth $\Gamma_{\rm opt}$ is much smaller than the original cavity bandwidth $\gamma$, and is tunable, principally through $\bar{G}_{0}$ which depends on the control light input power \cite{OMITweis}.

\begin{figure}[!t]
\includegraphics[width=0.38\textwidth]{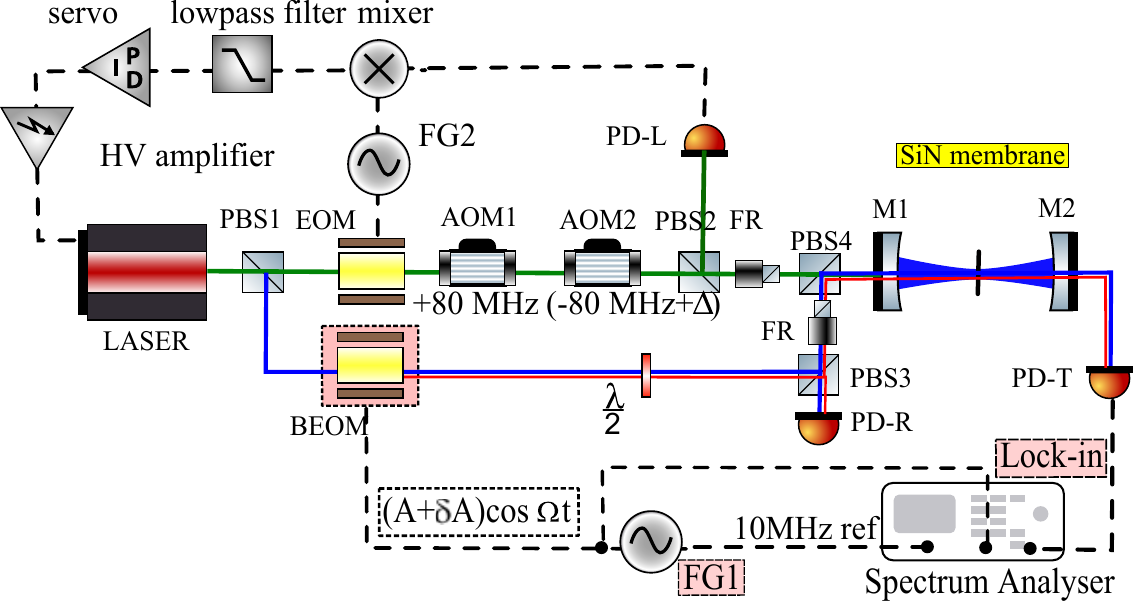}
\caption{(Color online)Experimental setup. The 85 mm long cavity sits in a vacuum chamber with a central silicon nitride membrane oscillator(1mm$\times$1mm$\times$50nm, effective mass $40$ ng). The green line represents the locking light for stabilizing the laser frequency to the cavity resonance using Pound-Drever-Hall (PDH) method \cite{pdh}. The blue line represents the control light, with  polarization orthogonal to the locking light. The broadband electro-optic modulator (BEOM) generates an upper-sideband from the control light, which is our signal light (red line). The $\Delta\sim400$ kHz  for the control light was created using a pair of $80$ MHz AOMs in the locking path.}
\label{fig:setup}
\end{figure}

When we inject the signal light into the system at different frequencies, we will observe a
rotation of the noise ellipse at the cavity output. For the single mode signal field with classical squeezed added noise, the rotation angle $\theta(\Omega)$ is determined by the phase response $\phi(\Omega)$ of the
cavity transmissivity $t(\Omega)$, which is given by:
\be
\theta(\Omega)=-\arctan{\frac{\Gamma_{\rm opt}(\Omega-\omega_{m})}{(\Omega-\omega_{m})^{2}+\gamma_{m}\Gamma_{\rm opt}}}
\ee
which is shown as theoretical curves in Fig. \ref{phaserotation} (b)(c).


\begin{figure}[!b]
\includegraphics[width=0.35\textwidth]{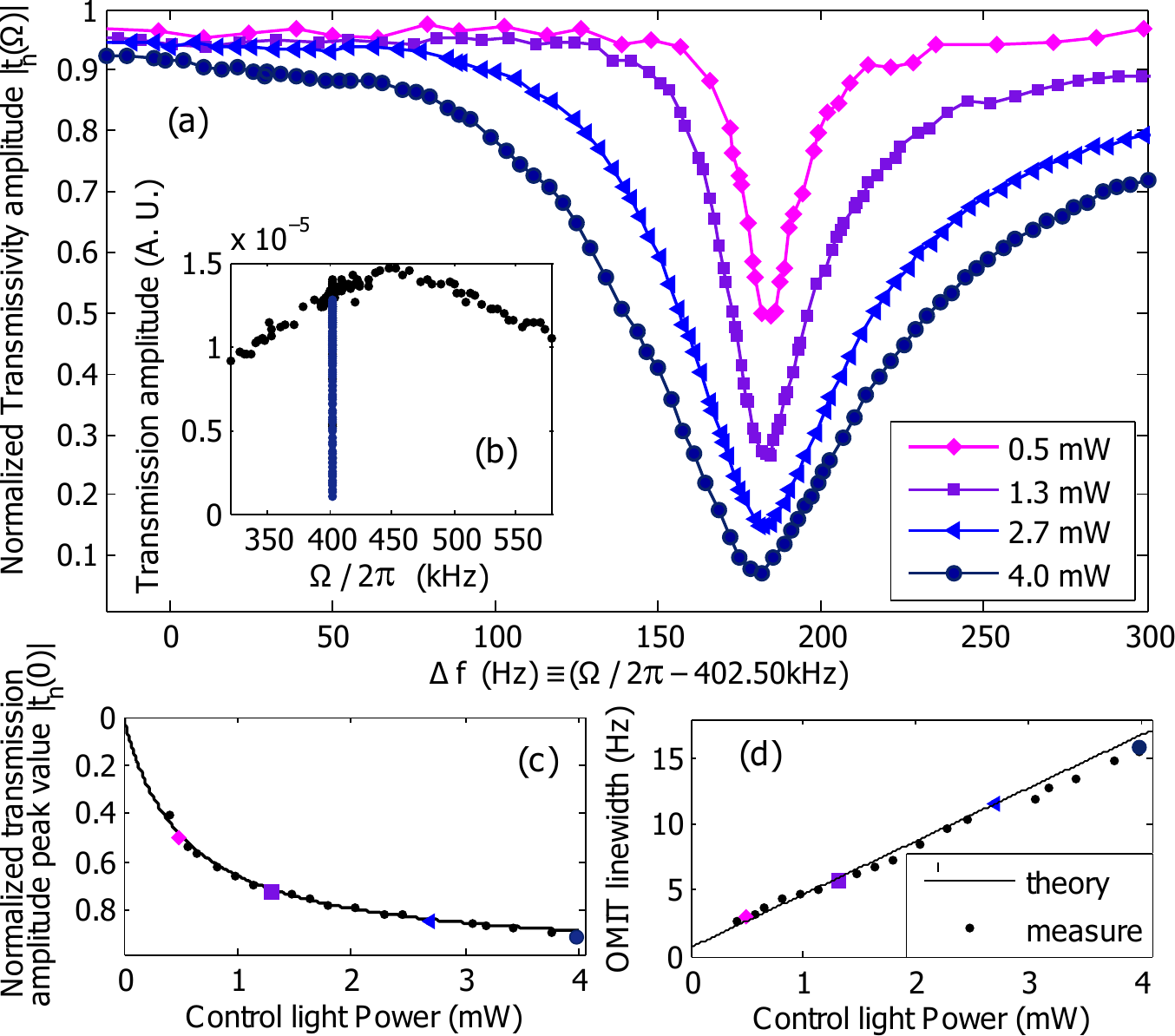}
\caption{Detected OMIT transmissivity. (a)Normalized transmissivity amplitude $|t_{n}(\Omega)|$ \textrm{vs}. frequency offset $\Delta f$ (Hz), where $\Delta f\equiv\Omega/2\pi-402.5$kHz. The control light powers were 0.5, 1.3, 2.7 and 4.0 mW respectively. (b) Transmissivity amplitude \textrm{vs}. frequency difference $\Omega/2\pi$ (kHz) in a span of 200kHz. The coupled cavity full linewidth was $170$ kHz in this measurement. (c) Normalized transmissivity amplitude peak value $|t_{n}(\Omega-\omega_{m} =0)|$ \textrm{vs}. the control light input power. (d) OMIT cavity full linewidth \textrm{vs}. the control light input power. The full linewidth data correspond to the Lorentzian transmissivity of the OMIT cavity. In this measurement, the mechanical resonance frequency was $\omega_{m}=2\pi\times402.7$ kHz. }
\label{fig:result}
\end{figure}

{\it Experimental Scheme---} In our experimental setup shown in Fig. \ref{fig:setup}, weak signal light is generated by passing the carrier control light through a broadband electro-optic modulator (BEOM). The BEOM modulates the control light and generates an upper-sideband, which is our signal field. The lower sideband $(\omega_p-\Omega)$ generated by the BEOM is far detuned from the cavity resonance, so it is totally reflected and can only be neglected at the transmission port. This method guarantees a common optical path for the signal light and the control light so as to avoid the fluctuating phase difference from an unlocked optical path. The voltage $V=A\cos{\Omega t}$ from function generator FG1, acting on the BEOM with modulation index $\beta=15$ mrad$/$V, determines the amplitude $|\hat a^s_{\rm in}|=\beta A |\hat a^c_{\rm in}|$ and the frequency $\omega_{s}=\omega_{p}+\Omega$ of the signal light. By adding random noise $\delta A$ to the voltage amplitude $A$, we increase the amplitude uncertainty of the signal light to simulate the ``phase squeezed light''.

Our optical cavity was mounted on an invar spacer machined by electrical discharge machining with accuracy of 0.1 $\mu \rm m$ and fixed in a vibration isolated vacuum tank. The M1 and M2 were clamped at the ends of the spacer. In order to optimize the optical coupling, we built an over-coupled cavity. The transmissivity $T1$ of M1 was chosen to be much larger than that of M2 ($T_{1}= 245.1\pm2.8$ ppm, $T_{2}=16.93\pm0.20$ ppm). This experiment was conducted at room temperature using a 1064nm Nd:YAG laser.

The mechanical oscillator in this study was a high $Q-$factor stoichiometric silicon nitride membrane window. In order to adjust the position and alignment of the membrane in the vacuum, it was attached to a piezoelectric actuator which was glued to a motorized optical mounts attached to the invar cavity spacer. To reduce the bonding loss, the membrane frame was bonded onto the actuator with Yacca gum, a natural resin with low intrinsic loss \cite{yacca}. After gluing, we measured the quality factor of the membrane with a He-Ne laser to characterize the extra mechanical loss $\gamma_{\rm gas}$ introduced by the background gas. When the background gas pressure $P_{\rm gas}$ is smaller than $3\times 10^{-5}$ mbar, the gas damping was negligibly small and the membrane quality factor was $\sim1.5\times 10^{6}$ at its mechanical resonance $\sim400$ kHz. (cf. Appendix A in supplemental material \cite{sup})

\begin{figure}[!b]
\includegraphics[width=0.38\textwidth]{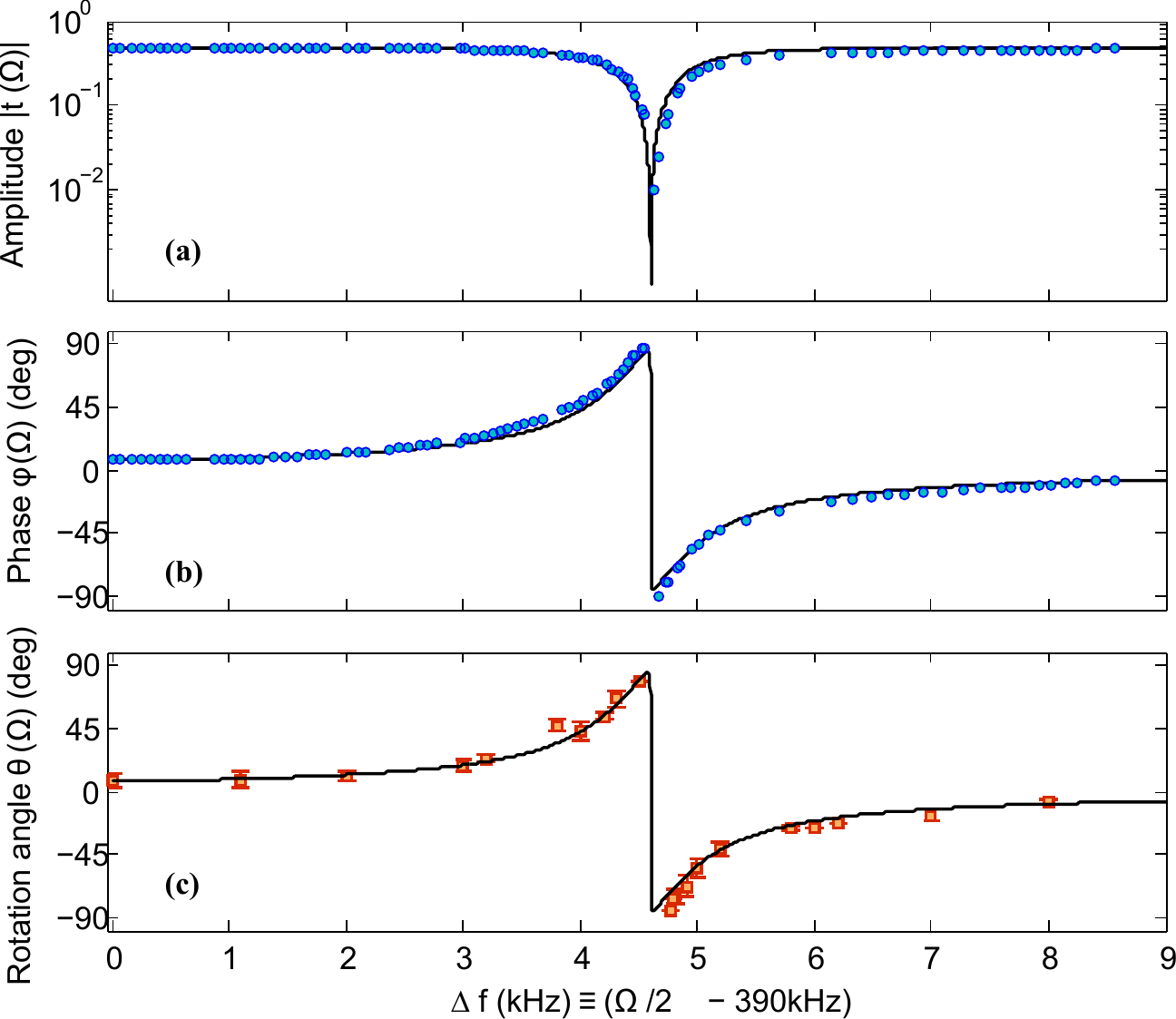}
\caption {OMIT cavity transmissivity amplitude $|t(\Omega)|$ (a), phase $\phi(\Omega)$ (b) and rotation angle $\theta(\Omega)$ (c) of the noise ellipse as a function of $\Delta f$ (kHz). The frequency offset is defined as $\Delta f\equiv\Omega/2\pi-390$kHz. In this measurement, the mechanical resonance frequency was $\omega_{m}=2\pi\times394.6$ kHz. }
\label{phaserotation}
\end{figure}

{\it Optomechanically induced transparency---}In our system, the linewidth of the OMIT cavity can be changed in two ways: (a) The input power of the control light can be adjusted by a half-wave plate before a polarized beam splitter (PBS3 in Fig. \ref{fig:setup}). (b) The coupled cavity linewidth $\gamma$ and the optomechanical coupling strength $G_0$ can be tuned by changing the position and alignment of the membrane in the cavity  \cite{coupling, finesse}(cf. Appendix B in \cite{sup}). We achieved a widely tunable linewidth of the OMIT cavity changing from 3 Hz to several hundred Hz.

In order to achieve an extremely narrow linewidth, we tuned the position and alignment of membrane and reduced the control light input power until the characteristic frequency $\Gamma_{\rm opt}$ was close to the mechanical linewidth $\gamma_m$. In Fig. \ref{fig:result}, we show the experimental results for the lowest linewidth data 3$\sim$15 Hz. Here, we define an normalized transmissivity amplitude as $t_{n}(\Omega)\equiv t(\Omega)/t_{0}$, where $t_{0}$ is the transmissivity amplitude of the signal light in the absence of the control field. The measurement data points in Fig. \ref{fig:result} (c) and (d) are well matched with the theoretical model shown as the black solid lines.

  {\it Frequency dependent noise ellipse rotation---} The above results show that OMIT effect can be used to create cavities with tunable linewidth down to a few Hz. We now demonstrate that such cavities have the appropriate phase response, and that they rotate the angle of the noise ellipse of the signal light as required for one simple filter cavity.
\begin{figure}[t]
\includegraphics[width=0.3\textwidth]{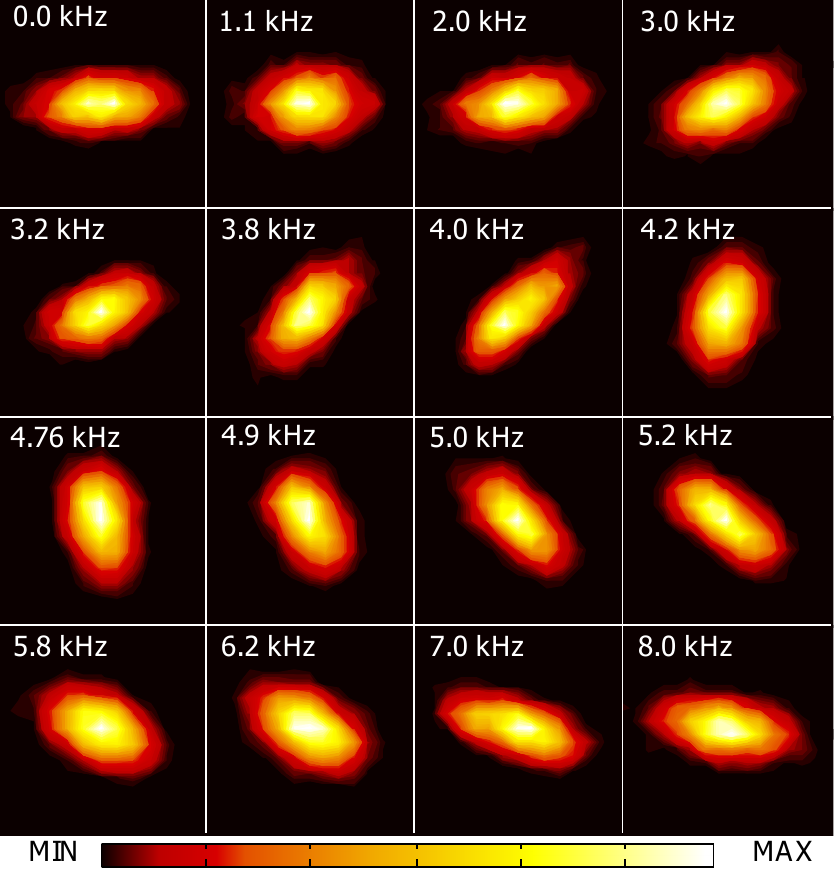}
\caption {Contour plots of the normalized noise ellipse in phasor diagram at different frequency offsets  $\Delta f$ corresponding to (c) in Fig. \ref{phaserotation}. In each phasor diagram, the horizontal axis is the amplitude quadrature and the vertical axis is the phase quadrature. The frequency offset of the OMIT cavity resonance is 4.6 kHz. As the frequency offset increases, the rotation pattern is changed from $0^{\circ}$ (off resonance) to $90^{\circ}$ anticlockwise. Near the resonance, it flips by $-180^{\circ}$. Above resonance, the rotation pattern is changed from $-90^{\circ}$ to $0^{\circ}$ anticlockwise.}
\label{rotate}
\end{figure}

 In order to characterize the noise ellipse rotation of the signal light in phasor diagram, we tuned the OMIT cavity linewidth to several hundred Hz and demonstrate the noise ellipse rotation in phasor diagram. The phase response and the rotation angles of the noise ellipses are detected by lock-in technique (See Fig. \ref{fig:setup} and \cite{EIT}). We take results of the coupled caivity with linewidth of $\sim 600$ Hz as an example.

In Fig. \ref{phaserotation} (a) and (b), we show the experimental results of the amplitude $|t(\Omega)|$ and the phase $\phi(\Omega)$ of the OMIT cavity transmission. The phase drop in the vicinity of the OMIT cavity ($|\Omega-\omega_m|<\sqrt{\gamma_m\Gamma_{\rm opt}}$) resonance was measured and shown in Appendix C \cite{sup}.

In Fig. \ref{phaserotation} (c) and Fig. \ref{rotate}, we show the measured rotation angles $\theta(\Omega)$ and the corresponding normalized noise ellipses in phasor diagram \cite{quantumnoise}. As shown in Fig. \ref{phaserotation} (c), the measured results for angle rotation of the noise ellipses well match both the theoretical model and the previous measurement of the phase $\phi(\Omega)$.

{\it Conclusion---}  We have shown an extremely narrow cavity linewidth created by optomechanical interaction in an optical cavity with a silicon nitride membrane in the middle.  Classical light with a noise ellipse simulating quantum squeezed light was injected into the cavity.  It demonstrates the frequency dependent noise ellipse rotation. The rotation angle follows the theoretical prediction in the detection band of advanced gravitational wave detectors. To use the current setup to develop a system for realizing frequency dependent squeezed vacuum in GW detectors in the future, it will be necessary to cool the resonator to the mK temperature range and dilute the mechanical losses by optical springs as discussed in references \cite{Yiqiu} and \cite{korth}.

We thank Haixing Miao, Yanbei Chen, Lisa Barsotti, Stefan Danilishin and D. E. McClelland for useful discussions. We thank the technicians Gary Light and Mark Dickinson for technical supports. We would like to thank the LIGO Scientific Collaboration, Gingin International Advisory, and our collaborators, especially Stefan Gosler, Gregg Harry, Bill Kells, Pierre-Francois Cohadon and Antoine Heidmann for useful advice. This research was supported by the Australian Research Council (DP120104676 and DP120100898).


\end{document}



\newcommand{\be}{\begin{equation}}
\newcommand{\ee}{\end{equation}}
\newcommand{\R}[1]{\textcolor{red}{#1}}
\newcommand{\B}[1]{\textcolor{blue}{#1}}


\title{Supplemental Material:\\
Classical demonstration of frequency dependent noise ellipse rotation using Optomechanically Induced Transparency}

\author{Jiayi Qin}
\email{jiayiqinphysics@gmail.com}
\affiliation{School of Physics, University of Western Australia, WA 6009, Australia}
\author{Chunnong Zhao}
\email{chunnong.zhao@uwa.edu.au}
\affiliation{School of Physics, University of Western Australia, WA 6009, Australia}
\author{Yiqiu Ma}
\email{myqphy@gmail.com}
\affiliation{School of Physics, University of Western Australia, WA 6009, Australia}
\author{Xu Chen}
\affiliation{School of Physics, University of Western Australia, WA 6009, Australia}
\author{Li Ju}
\affiliation{School of Physics, University of Western Australia, WA 6009, Australia}
\author{David G. Blair}
\affiliation{School of Physics, University of Western Australia, WA 6009, Australia}

\maketitle
\section{APPENDIX}
\subsection{A: Mechanical linewidth}\label{AppA}
When the background gas pressure $P_{\rm gas}>10^{-2}$ mbar, the size of the membrane is ignorable compared with the mean free path of the background gas. This satisfies the requirements for the non-interacting gas molecule model \cite{q}. The background gas leads to a mean damping force $dp/dt=-\gamma_{\rm gas}p/2$. The background gas damping rate is $\gamma_{\rm gas}=16 P_{\rm gas}/ (\pi \rho_{m} d \bar{v}_{\rm gas})$, where $d=50$ nm is the membrane thickness and $\bar{v}_{\rm gas}=\sqrt{3 R T /M_{\rm gas}}$ is the gas mean speed \cite{q2}. The quality factor is given by $Q=2\pi \omega_{m}/(\gamma_{m}+\gamma_{\rm gas})$.

\begin{figure}[!h]
\includegraphics[width=0.35\textwidth]{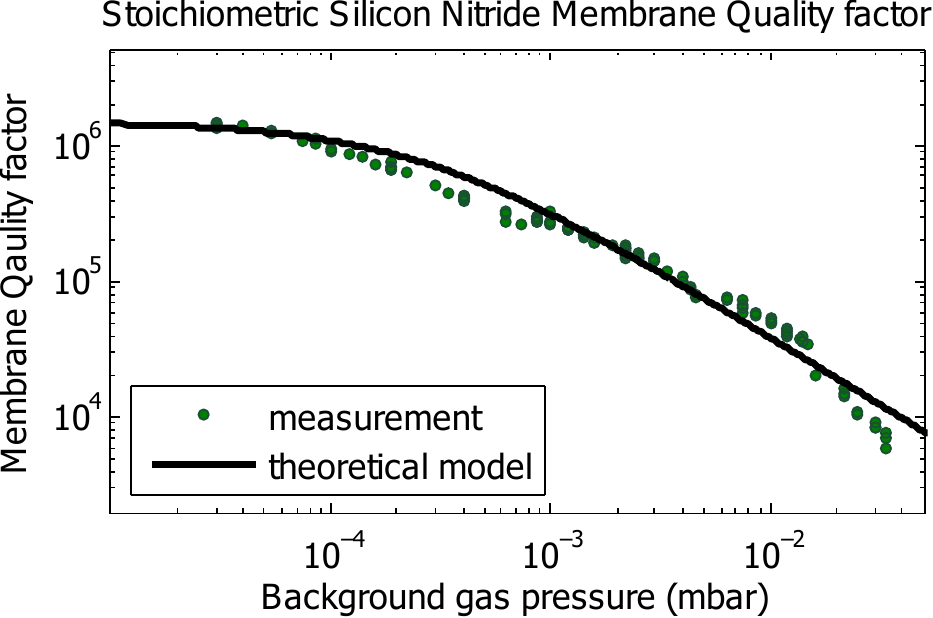}
\caption {Measured membrane quality factor vs background gas pressure.}
\label{q}
\end{figure}

We used He-Ne laser to measure the ringdown time $\tau$ of the mechanical oscillation amplitude in different pressures. The quality factor is given by $Q=\pi f\tau$. As shown in Fig. \ref{q}, when the background gas pressure $P_{\rm gas}$ reach the order of $10^{-5}$ mbar, $\gamma_{\rm gas}$ is negligible small and the quality factor of the membrane is $\sim1.5\times10^{6}$ at its resonant frequency of $\sim400$ kHz.

\subsection{B: Coupled cavity linewidth $\gamma$ and optomechanical coupling constant $G_{0}$ }
\begin{figure}[!h]
\includegraphics[width=0.38\textwidth]{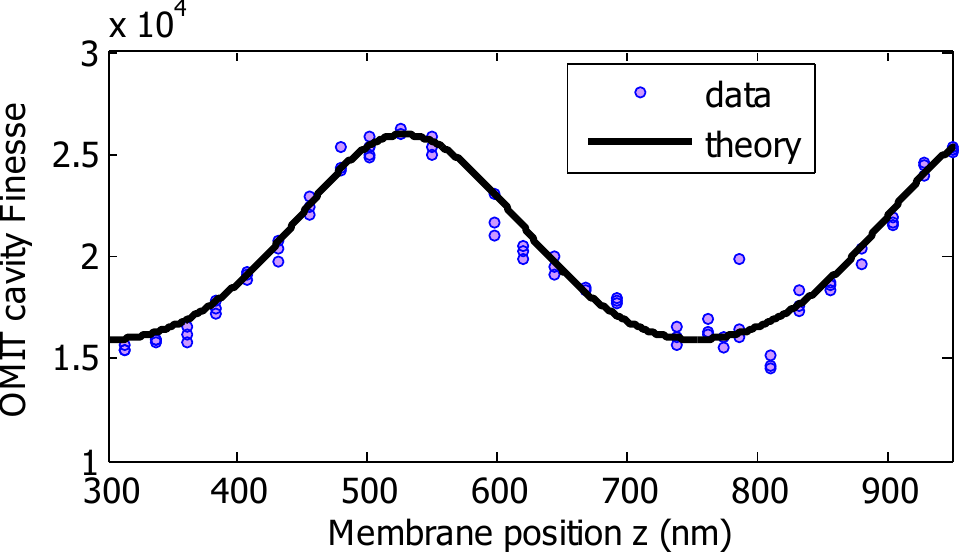}
\caption {Coupled cavity finesse $\mathcal{F}(z)$ vs. membrane position $z$. Optical wave length is $\lambda=1064$ nm. The empty cavity finesse is $\mathcal{F}_{0}=22400$. The membrane index of refraction $n=2+ 2.5\times10^{-5}i$.}
\label{finesse}
\end{figure}

In the coupled cavity system, optical absorption by the membrane changes the coupled cavity
 $\mathcal{F}(z)$ in different positions along the cavity axis \cite{coupling, finesse}. The coupled cavity linewidth $\gamma=\pi f_{\rm FSR}/\mathcal{F}$, where $f_{\rm FSR}=c/2L$ is the free spectral range of the cavity. According to previous theoretical work \cite{finesse}, the coupled cavity finesse $\mathcal{F}(z)$ is a periodic function of the membrane position $z$ with period equal to half wavelength of the optical mode. In Fig. \ref{finesse}, we show the experimental results of the coupled cavity finesse $\mathcal{F}(z)$ as a function of $z$. And the optical loss of the system also depends on the alignment of the membrane in the transverse plane orthogonal to the cavity axis.

The linear optomechanical coupling constant $G_{0}$ is the change of cavity resonance frequency per unit displacement of the membrane. In our setup, we have:\cite{coupling}
\be\label{G0}
G_{0}=\partial_{z}\omega_{c}(z)=\partial_{z}[\frac{c}{L}\arccos(|r_{m}|\cos\frac{4\pi z}{\lambda})],
\ee
where $r_{m}$ is the electronic field reflectivity of the membrane. Therefore, the coupling constant $G_{0}$ depends on the center-of-mass position along the cavity axis.

\subsection{C: Phase response near the OMIT cavity resonance}
\begin{figure}[!h]
\includegraphics[width=0.38\textwidth]{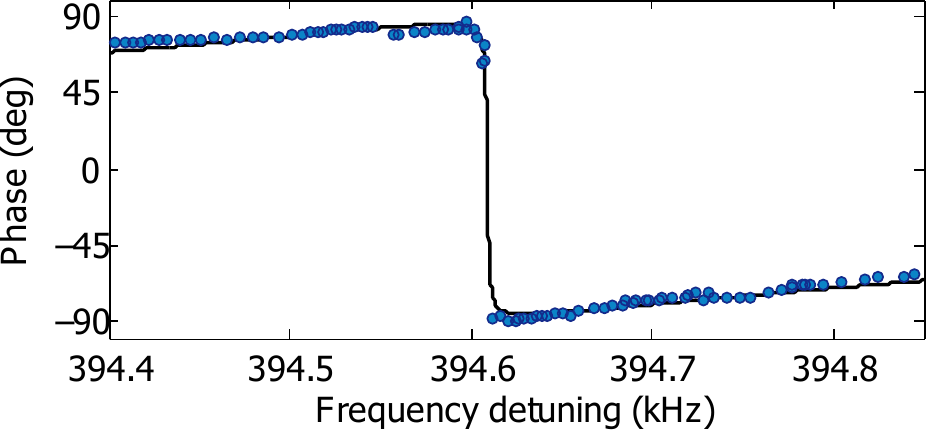}
\caption {The phase $\phi(\Omega)$ of the OMIT cavity transmission. The minimum sweep step was 0.25 Hz and the intensity of the signal on the resonance was very small.}
\label{fig:400hz}
\end{figure}
